\documentclass[12pt,preprint]{aastex}
\shorttitle{Phosphorus Oscillator Strengths}
\shortauthors{Federman et al.}

\begin{document}
\title{Oscillator Strengths for Ultraviolet Transitions in 
P {\small II}}
\author{S.R. Federman, M. Brown, S. Torok, S. Cheng, 
R.E. Irving, R.M. Schectman, and L.J. Curtis}
\affil{Department of Physics and Astronomy, University of Toledo, 
Toledo, OH 43606}
\email{steven.federman@utoledo.edu; mbrown@physics.utoledo.edu; 
storok@physics.utoledo.edu; scheng@physics.utoledo.edu; 
rirving@physics.utoledo.edu; rms@physics.utoledo.edu; 
ljc@physics.utoledo.edu.}

\begin{abstract}

We report lifetimes, branching fractions, and the resulting 
oscillator strengths for transitions within the P~{\small II} 
multiplet (3$s^2$3$p^2$ $^3P$ $-$ 3$s^2$3$p$4$s$ $^3P^o$) at 
1154 \AA.  These beam-foil measurements represent the most 
comprehensive and precise set currently available experimentally.  
Comparison with earlier experimental and theoretical results 
is very good.  Since Morton's most recent compilation 
is based on the earlier body of results, phosphorus abundances 
for interstellar material in our Galaxy and beyond derived from 
$\lambda$1154 do not require any revision.

\end{abstract}


\section{Introduction}

Singly-ionized phosphorus is the dominant form of this element 
in the neutral interstellar medium.  
Most astronomical studies rely on the 
$^3P_0$ $-$ $^3P^o_1$ transition at 1153 \AA\ because it is 
one of the strongest in P~{\small II} and is in a relatively 
clean portion of the spectrum.  For instance, observations of 
diffuse clouds in the Galaxy (Dufton, Keenan, \& Hibbert 
1986; Jenkins, Savage, \& Spitzer 1986) and in the Small 
Magellanic Cloud (Mallouris et al. 2001) focused on the amount 
of phosphorus depleted onto the surface of interstellar 
grains.  In more distant galaxies and damped Lyman-$\alpha$ 
systems, P~{\small II} absorption then reveals the metallicity 
of the gas and the system's nucleosynthetic history (e.g., 
Molaro et al. 2001; Levshakov et al. 2002; Pettini et al. 2002).  
Here we present the most comprehensive and precise experimental 
results to date for all transitions within the multiplet 
associated with the line $\lambda$1153.

The above astronomical studies use oscillator strengths 
($f$-values) compiled by Morton (1991, 2003) from the 
theoretical computations of Hibbert (1988) and experimental 
lifetime of Livingston et al. (1975).  Hibbert's results 
represent an improvement over his earlier calculation (Hibbert 
1986), which did not include the 3$s^2$3$p$4$p$ states.  Other 
work on the 3$s^2$3$p^2$ $^3P$ $-$ 3$s^2$3$p$4$s$ $^3P^o$ 
multiplet includes an experiment (Smith 1978) and computational 
efforts (Brage, Merkelis, \& Froese Fischer 1993; Tayal 2003; 
Froese Fischer, Tachiev, \& Irimia 2006).  
Our beam-foil measurements on lifetimes and branching fractions 
allow us to derive the first purely experimental $f$-values for 
comparison with theory.  Our results are presented and then 
discussed after we describe the experiment.

\section{Experimental Details}

The beam-foil measurements were performed at the Toledo Heavy 
Ion Accelerator.  Details about the facility and general 
aspects of the experimental procedures can be found in our 
earlier papers (e.g., Federman et al. 1992; Haar et al. 1993; 
Schectman et al. 2000).  Here we focus on the particulars for 
the P~{\small II} data.  Phosphorus ions were produced by 
heating red phosphorus in a low-temperature oven and by 
subsequent charging through interactions in an Ar plasma.  
Most lifetime measurements and the 
spectra for branching fractions were obtained at an energy of 
170 keV.  Systematic effects, such as beam divergence, foil 
thickening, and nuclear scattering, were studied through 
comparisons of forward and reverse decay curves and of data 
acquired at 240 keV.  Typical P$^+$ beam currents were 
200 nA.  The ions emerged in a variety of charge states and 
excited states upon passage through carbon foils whose 
thicknesses were 2.4 $\mu$g cm$^{-2}$.  An Acton 1 m 
normal-incidence vacuum ultraviolet monochromator with a 
2400 line mm$^{-1}$ grating blazed at 800 \AA\ was used 
for the transitions involving the multiplet at 1154 \AA.  
The radiation was detected with a Galileo channeltron electron 
multiplier.  Cascades that could repopulate the upper states 
of interest with wavelengths in the visible portion of the 
spectrum were sought with a 600 line mm$^{-1}$ grating 
blazed at 3000 \AA\ and an S20 photomultiplier (Centronic 
Q4283) cooled with dry ice.

Lifetimes for the three values of $J$ in the 
upper fine structure levels of the multiplet $\lambda$1154 
were obtained from decay curves for all transitions 
between 3$p^2$ $^3P$ and 3$p$4$s$ $^3P^o$.  The lifetimes 
were extracted from multiexponent fits.  
The decay curves were best fit by 
two exponentials; we ascribe the weaker and longer-lived 
decay to a cascading transition that repopulates the 
upper level of interest.  Figure 1 shows the decay curve 
for the $^3P_2$ $-$ $^3P^o_2$ line at 1154 \AA.  The presence 
of the second decay necessitated our acquiring measurements 
on all six transitions within the multiplet for the 
derivation of accurate branching fractions.  This yielded
more precise determinations for the primary decay 
from the upper levels with $J$ $=$ 1 and 2.

A search for the transitions at visible wavelengths that 
were responsible for repopulating $^3P^o$ was conducted.  
The purpose was to obtain an independent measure of the 
lifetime in order to perform the method of arbitrarily 
normalized decay curves (ANDC) (see Curtis, 
Berry, \& Bromander 1971).  The 
ANDC method leads to a more secure value for the lifetime 
of the primary decay of most interest to the current 
study.  Our measurements revealed a lifetime for the 
secondary decay of about 10 ns.  Large-scale computations 
(Hibbert 1988; Tayal 2003; Froese Fischer et al. 
2006) suggest that transitions 
originating from 3$p$4$p$ $^3D$ cause the repopulation.  
The lines occur around 6000 \AA, but were too weak to 
enable us to carry out the analysis.

Branching fractions are needed to convert lifetimes 
into oscillator strengths when more than one channel for 
decay is present.  Theoretical (Hibbert 1988; Tayal 
2003; Froese Fischer et al. 2006) 
and semi-empirical (Curtis 2000) calculations 
indicate that intercombination lines arising from the 
3$p$4$s$ $^3P^o$ state are weak (branching fractions 
less than 2\%, which are below the sensitivity of our 
experiment).  As a result, we focused on branching among 
the dipole-allowed transitions between 1150 and 1160 \AA; 
the scan revealing these transitions, and from which 
branching fractions were obtained, appears in Fig. 2.  
This was accomplished by determining the relative 
integrated intensities from lines with a common upper 
level through Gaussian fits.  Since the fits indicated 
line widths that were indistinguishable from one another, 
we relied on the intensities, after correcting for the 
contribution (about 10\%) from the secondary decay noted 
above.  Because the spectral interval is small, 
systematic differences in instrumental response are 
not discernible with our sensitivity.  This point is 
addressed further below.

\section{Results and Discussion}

The results of our lifetime measurements appear in Table 1 
and the oscillator strengths derived from our branching 
fractions are given in Table 2.  The tables also provide 
comparisons with earlier experimental (Livingston et al. 
1975; Smith 1978) and theoretical work (Hibbert 1988; 
Brage et al. 1993; Tayal 2003; Froese Fischer et al. 
2006).  The preferred $f$-values 
from Morton's most recent compilation (2003), which are 
based on the work of Livingston et al. and Hibbert, are 
included in Table 2.  The experimentally determined 
branching fractions used to transform the lifetimes 
for the $J_u$ $=$ 1 level into $f$-values are 
$0.359 \pm 0.027$ ($J_l$ $=$ 0), $0.254 \pm 0.014$ 
($J_l$ $=$ 1), and $0.387 \pm 0.026$ ($J_l$ $=$ 2).  The 
corresponding values for the $J_u$ $=$ 2 level are 
$0.267 \pm 0.016$ ($J_l$ $=$ 1) and $0.733 \pm 0.033$ 
($J_l$ $=$ 2).  Our branching fractions agree very well 
with those derived from theoretical and semi-empirical 
analyses (Hibbert 1988; Brage et al. 1993; Curtis 2000; 
Tayal 2003; Froese Fischer et al. 2006), 
as can be inferred from the comparison in 
Table 2.

The agreement between our results for 
lifetimes and $f$-values and others is very good.  
Although our determinations are the most precise ones 
currently available experimentally, the agreement with 
theory has not suffered.  This is not unexpected, since 
the intercombination lines involving 3$p$4$s$ $^3P^o$ 
as the upper state are quite weak and there is little 
configuration interaction (CI) involving the upper or lower 
levels of the multiplet.  On the other hand, CI is much 
more prominent in other P~{\small II} multiplets, including 
$\lambda\lambda$964, 967, and 1308 
(respectively 3$s^2$3$p^2$ $^3P$ $-$ 3$s^2$3$p$3$d$ $^3D^o$, 
3$s^2$3$p$3$d$ $^3P^o$, 3$s$3$p^3$ $^3P^o$), 
and the agreement among 
theoretical calculations is much worse.  Since astronomical 
studies make use of transitions in these three multiplets, 
we plan future measurements in an attempt to resolve the 
discrepancies now present.

We also were able to obtain the lifetime 
for each upper fine structure level 
and branching fractions for all transitions within the 
multiplet.  Thus, our measurements are the most 
comprehensive to date.  This set of data can be used to 
test the hypothesis (e.g., Curtis 2000) that in cases where 
there is little CI, semi-empirically derived branching 
fractions can provide a means to help calibrate 
instruments below 2000 \AA.  This will be discussed further 
elsewhere.

\section{Conclusions}

We presented beam-foil measurements on lifetimes and 
branching fractions for the multiplet at 1154 \AA\ in 
P~{\small II}.  These were combined to yield $f$-values 
that are needed for studies of interstellar abundances and 
the history of nucleosynthesis from quasar absorption-line 
systems.  The measurements represent the most comprehensive 
and precise set available experimentally for the 
multiplet.  The lifetimes and oscillator strengths agree 
very well with earlier determinations and Morton's (2003) 
compilation.  Since the intercombination lines arising 
from the upper levels in the multiplet are weak, and 
since there is little configuration interaction, the close 
agreement is not unanticipated.  There are 
transitions in other multiplets (at 964, 967, and 1308 
\AA) that are used in the astronomical studies.  
Configuration interaction is much stronger for these 
three multiplets and the available results differ 
significantly; we plan future measurements to help 
resolve the discrepancies.  Then phosphorus abundances 
based on any of the well-studied ultraviolet lines will 
be known more securely.

\acknowledgments
This work was supported by NASA grants NAG5-11440 and NNG06GC70G.  
S.T. acknowledges support by the National Science Foundation under 
Grant No. 0353899.

\begin{deluxetable}{cccccccc}
%\rotate
\tablecolumns{8}
\tablewidth{0pt}
\tabletypesize{\scriptsize}
\tablecaption{P~{\small II} Lifetimes for 3$s^2$3$p$4$s$ 
$^3P^o$ Levels}
\startdata
\hline \hline\\
$J_u$ & \multicolumn{7}{c}{$\tau$ (ns)} 
\\ \cline{2-8}
& Present & LKIP\tablenotemark{a} & S\tablenotemark{b} & 
H\tablenotemark{c} & BMF\tablenotemark{d} & T\tablenotemark{e} 
& FTI\tablenotemark{f} \\ \hline
0 & $0.79 \pm 0.10$ & $\ldots$ & $\ldots$ & 0.82 & $\ldots$ & 0.784 
& 0.796 \\
1 & $0.79 \pm 0.06$ & $\ldots$ & $\ldots$ & 0.81 & $\ldots$ & 0.778 
& 0.789 \\
2 & $0.84 \pm 0.07$ & $0.85 \pm 0.11$ & $1.3 \pm 0.5$ & 0.80 & 0.80 & 
0.772 & 0.776 \\
\enddata
\tablenotetext{a}{Livingston et al. 1975 $-$ beam-foil experiment.}
\tablenotetext{b}{Smith 1978 $-$ phase shift experiment.}
\tablenotetext{c}{Hibbert 1988 $-$ configuration interaction calculation.}
\tablenotetext{d}{Brage et al. 1993 $-$ multi-configuration Hartree-Fock 
calculation.}
\tablenotetext{e}{Tayal 2003 $-$ multi-configuration Hartree-Fock 
calculation.}
\tablenotetext{f}{Froese Fischer et al. 2006 $-$ multi-configuration 
Hartree-Fock calculation.}
\end{deluxetable}

\begin{deluxetable}{ccccccccccc}
%\rotate
\tablecolumns{11}
\tablewidth{0pt}
\tabletypesize{\scriptsize}
\tablecaption{P~{\small II} Oscillator Strengths for the Multiplet 
3$s^2$3$p^2$ $^3P$ $-$ 3$s^2$3$p$4$s$ $^3P^o$}
\startdata
\hline \hline\\
$\lambda_{ul}$ (\AA) & $J_l$ & $J_u$ & 
\multicolumn{7}{c}{$f$-value ($\times 10^{-2}$)} \\ \cline{4-11}
 & & & Present & LKIP\tablenotemark{a} & S\tablenotemark{b} & 
H\tablenotemark{c} & BMF\tablenotemark{d} & T\tablenotemark{e} & 
FTI\tablenotemark{f} & M\tablenotemark{g} \\ \hline
1149.958 & 1 & 2 & $10.5 \pm 1.1$ & $\ldots$ & 
$\ldots$ & 10.4\tablenotemark{h} & $\ldots$ & 10.8\tablenotemark{h} & 
10.8 & 10.4 \\
 & $\ldots$ & $\ldots$ & $\ldots$ & $\ldots$ & $\ldots$ &
11.3\tablenotemark{i} & $\ldots$ & 10.4\tablenotemark{i} & 
$\ldots$ & $\ldots$ \\
1152.818 & 0 & 1 & $27.2 \pm 2.9$ & $\ldots$ &
$\ldots$ & 24.4\tablenotemark{h} & $\ldots$ & 25.1\tablenotemark{h} & 
25.3 & 24.5 \\
 & $\ldots$ & $\ldots$ & $\ldots$ & $\ldots$ & $\ldots$ &
26.4\tablenotemark{i} & $\ldots$ & 24.1\tablenotemark{i} & 
$\ldots$ & $\ldots$ \\
1153.995 & 2 & 2 & $17.4 \pm 1.6$ & $\ldots$ &
$\ldots$ & 18.6\tablenotemark{h} & $\ldots$ & 19.2\tablenotemark{h} & 
19.3 & 18.6 \\
 & $\ldots$ & $\ldots$ & $\ldots$ & $\ldots$ & $\ldots$ &
20.1\tablenotemark{i} & $\ldots$ & 18.4\tablenotemark{i} & 
$\ldots$ & $\ldots$ \\
1155.014 & 1 & 1 & $6.4 \pm 0.6$ & $\ldots$ &
$\ldots$ & 6.1\tablenotemark{h} & $\ldots$ & 6.2\tablenotemark{h} & 
6.3 & 6.1 \\
 & $\ldots$ & $\ldots$ & $\ldots$ & $\ldots$ & $\ldots$ &
6.5\tablenotemark{i} & $\ldots$ & 6.0\tablenotemark{i} & 
$\ldots$ & $\ldots$ \\
1156.970 & 1 & 0 & $8.5 \pm 1.1$ & $\ldots$ & 
$\ldots$ & 8.1\tablenotemark{h} & $\ldots$ & 8.5\tablenotemark{h} & 
8.4 & 8.2 \\
 & $\ldots$ & $\ldots$ & $\ldots$ & $\ldots$ & $\ldots$ &
8.8\tablenotemark{i} & $\ldots$ & 8.1\tablenotemark{i} & 
$\ldots$ & $\ldots$ \\
1159.086 & 2 & 1 & $5.9 \pm 0.6$ & $\ldots$ & 
$\ldots$ & 6.2\tablenotemark{h} & $\ldots$ & 6.3\tablenotemark{h} & 
6.3 & 6.2 \\
 & $\ldots$ & $\ldots$ & $\ldots$ & $\ldots$ & $\ldots$ &
6.7\tablenotemark{i} & $\ldots$ & 6.0\tablenotemark{i} & 
$\ldots$ & $\ldots$ \\
Multiplet & $\ldots$ & $\ldots$ & $24.4 \pm 1.2$ & $23 \pm 3$ & 15.0 & 
24.7\tablenotemark{h} & 25.2\tablenotemark{h} & 
$\ldots$ & $\ldots$ & 24.7 \\
 & $\ldots$ & $\ldots$ & $\ldots$ & $\ldots$ & $\ldots$ & 
26.7\tablenotemark{i} & 25.0\tablenotemark{i} & 
$\ldots$ & $\ldots$ & $\ldots$ \\
\enddata
\tablenotetext{a}{Livingston et al. 1975.}
\tablenotetext{b}{Smith 1978.}
\tablenotetext{c}{Hibbert 1988.}
\tablenotetext{d}{Brage et al. 1993.}
\tablenotetext{e}{Tayal 2003.}
\tablenotetext{f}{Froese Fischer et al. 2006.}
\tablenotetext{g}{Morton 2003 compilation.}
\tablenotetext{h}{Based on length formalism.}
\tablenotetext{i}{Based on velocity formalism.}
\end{deluxetable}

\clearpage

\begin{figure}
\begin{center}
\includegraphics[scale=1.0]{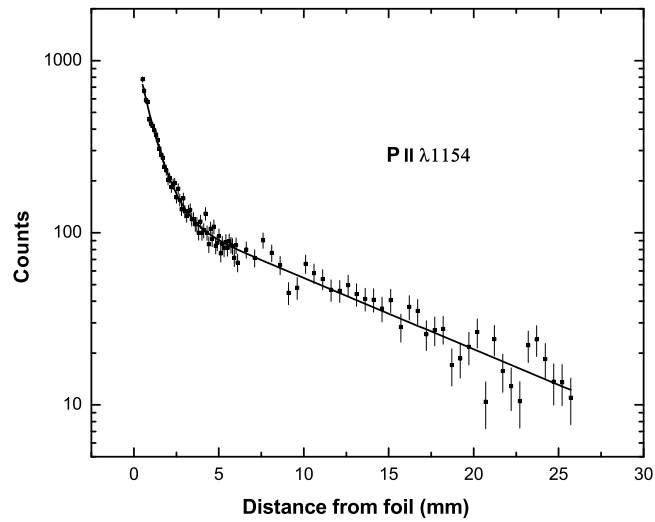}
\end{center}
\vspace{0.3in}
\caption{The measured P~{\small II} decay curve for the line at 1154 
\AA\ for a beam energy of 170 keV.  The post-foil beam velocity at
this energy was 1.0068 mm ns$^{-1}$, thus establishing the time since
excitation for a given foil position.  The foil was moved relative
to the monochromator entrance slit in increments of 0.1 mm until it
was displaced 5 mm; then the increments were increased to 0.5 mm.
A two-exponential fit to the data is shown by the solid curve.}
\end{figure}

\clearpage

\begin{figure}
\begin{center}
\includegraphics[scale=1.0]{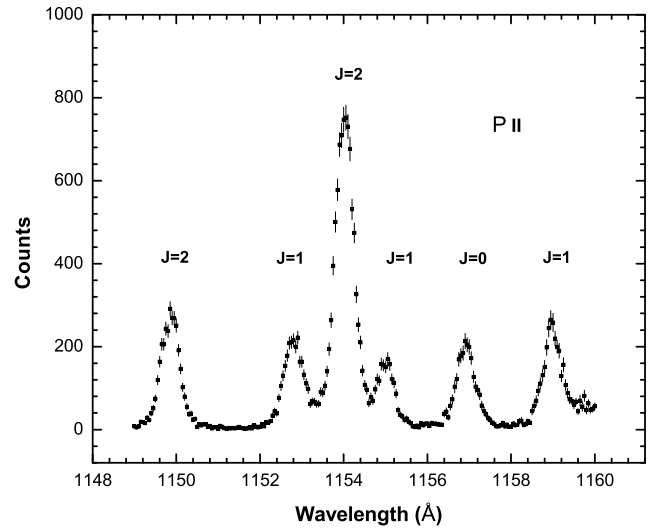}
\end{center}
\vspace{0.3in}
\caption{A spectrum of the P~{\small II} multiplet at 1154 \AA.  
The total angular momentum quantum number for each
upper fine structure level is indicated.}
\end{figure}

\end{document}